\documentclass[aps,prb,preprint,showpacs]{revtex4-1}

\usepackage{epsfig}
\usepackage{graphicx}
\usepackage[title]{appendix}
\usepackage{amssymb}
\usepackage{amsfonts}
\usepackage{amsmath}
\usepackage{amsbsy}
\usepackage{color}
\usepackage{epstopdf}
\usepackage{subfigure}
\usepackage{float}
\renewcommand{\vec}[1]{\mbox{\boldmath$\mathrm{#1}$}}
\let\sb=_ \catcode`\_=\active \def_#1{\ensuremath \sb{\rm#1}}

\begin{document}

\title{Gate-controlled magnon-assisted  switching of magnetization in ferroelectric/ferromagnetic junctions}

\author{Yaojin Li$^1$}

\author{Min Chen$^1$}

\author{Jamal Berakdar$^2$}

\author{Chenglong Jia$^{1,2}$}

\affiliation{$^1$Key Laboratory for Magnetism and Magnetic Materials of the Ministry of Education, Lanzhou University, Lanzhou 730000, China \\
$^2$Institut f\"ur Physik, Martin-Luther Universit\"at Halle-Wittenberg, Halle (Saale) 06099, Germany}

\begin{abstract}
Interfacing a ferromagnet with a polarized ferroelectric gate generates a non-uniform, interfacial spin density coupled to the ferroelectric polarization.  \textcolor{black}{This coupling allows
for an electric field control of the effective  field acting on the magnetization. To unravel the usefulness of this interfacial   magneto-electric coupling  we investigate
 the  magnetization dynamics of
 a ferroelectric/ferromagnetic  multilayer structure  using   the Landau-Lifshitz-Baryakhtar equation. The results  demonstrate that the interfacial magnetoelectric coupling is utilizable as a highly localized and efficient tool for manipulating magnetism by electrical means. Ways of enhancing the strength of the interfacial coupling and/or its effects are discussed.}

\end{abstract}

\pacs{75.78.-n, 77.55.Nv, 75.60.Jk}


\maketitle
\textcolor{black}{\section{Introduction}}

Electrical control of magnetism  has the potential to boost spintronic devices with a number of  novel functionalities \cite{Eerenstein:2006km,M. Weisheit:2007,T. Maruyama:2009,D. Chiba:2011,Jia:2015iz}. To mention an example,
magnetization switching can be achieved via a spin-polarized electric current due to the  spin-transfer torque or the spin-orbital torque in the presence of a spin orbital interaction \cite{J. C. Slonczewski:1996,L. Berger:1996,J. A. Katine:2000,M. D. Stiles:2002,Y. Tserkovnyak:2008,Brataas:2012fb,Fan:2014hb,Brataas:2014dla,Oh:2016ev,Fukami:2016kq}. One may also use an electric field  to manipulate the
 magnetization  dynamics \cite{Vaz:2012dp,T. Y. Liu:2011,T. Nozaki:2012,Brovko:2014gsb,Schueler2017,Matsukura:2015hya,Y. Shiota:2016} in which case the electric field
may lead to modulations in the charge carrier density or may affect the magnetic properties such as the magnetic moment, the exchange interaction and/or the magnetic anisotropy \cite{Vaz:2012dp,Brovko:2014gsb,Schueler2017,Matsukura:2015hya}.
Compared to driving  magnetization via a spin-polarized current, an  electric field governing the  magnetization has a clear advantage
as it allows for non-volatile device concepts with significantly reduced  energy dissipation.
On the other hand,  an external electric field applied to an itinerant ferromagnet (FM) is shielded by charge accumulation or depletion caused by spin-dependent screening charge that extends  on a length scale of only a few angstroms into the FM \cite{Zhang:1999cx}. This extreme surface confinement of screening  hinders its utilization to steer the magnetic dynamics of bulk or a relatively thick nanometer-sized FM \cite{Shiota:2012dh,Wang:2012jf}. Experimentally, ultra-thin metallic FM films were thus necessary to observe an electric field influence on the dynamic of an FM  \cite{Vaz:2012dp,Brovko:2014gsb,Nan:2014ck}.

In this work we show that  while the spin-polarized  screening charge is surface confined,  in the spin channel
a local non-uniform spiral spin density builds up at the interface and goes over into  the initial uniform (bulk)  magnetization away from the interface.
Hence, this  interfacial spin spiral acts as  a topological defect in the  initial uniform magnetization vector field. The range of the  spiral defect is set by the spin diffusion length $\lambda_m$ \cite{J. Bass:2007}
which is much larger than  the charge screening length.This spin-spiral constitutes a magnetoelectric effect that has a substantial influence on the traversal magnetization dynamics of
FM layers with thickness over tens of nanometers \cite{footnot}.
The interfacial spiral spin density can be viewed as a magnonic accumulation stabilized by the interfacial, spin-dependent charge rearrangement at  the contact region between the FM and the ferroelectrics (having the FE polarization $\vec{P}$)  and by the
uniform  (bulk) magnetization of FM far away from the interface \cite{Jia C. L:2014}.
 $\vec{P}$ responds to an external electric field and so does the magnetic dynamics.
As shown below,  this magnonic-assisted magnetoelectric coupling arising when using a dielectric FE gate,  allows a (ferro)electric field control of the effective driving  field
that governs the magnetization switching of  a FM layer with a thickness on the range of  the spin diffusion length $\lambda_m$, which is clearly of an advantage for designing spin-based, non-volatile  nanoelectronic devices.  \\

In Sec. \ref{sec1} we discuss the mathematical details of the spin-spiral magnetoelectric coupling, followed by its implementation into the equations of  motion for the
  magnetization dynamics  in Sec. \ref{sec2}. In Sec. \ref{sec3} results of numerical simulations are presented and discussed showing  to which extent  the spin-spiral magnetoelectric coupling  can allow for the electric field control of the magnetization in FE/FM composites. Ways to enhance the effects are discussed and brief conclusions are made in Sec. \ref{sec4}.

\textcolor{black}{\section{Interfacial magnetoelectric coupling}\label{sec1}}

Theoretically, the above magnon accumulation  scenario maybe viewed as follows:
\textcolor{black}{When a  FE layer with remanent  electric  polarization $\vec{P}$ and   surface charges $\sigma_{FE}$ is brought in contact with an itinerant (charge-neutral) FM,
bond rearrangements occur within few atomic layers in  the interface vicinity \cite{prl08andothers}.
On the FM side, the rearranged   spin polarized charge density  implies a spin configuration different from the bulk one. The modifications of the magnitude
of the interfacial  local magnetic moments are dictated by hybridization and charge transfer  and were studied thoroughly  both theoretically and experimentally
(e.g., Ref.[\onlinecite{prl08andothers}]). Here we are interested in the consequence on the long-range magnetic order extending to the asymptotic bulk magnetization.
 In the mean-field formulations, the induced spin density $\vec{s}$ is exchange coupled with the localized magnetic moments $\vec{S}$, which can be  treated classically as an effective magnetization $\vec{M} = -\frac{g\mu_{B}}{a^{3}} \vec{S}$ with $\mu_{B}$, $g$, and $a$ being the Bohr magneton, g-factor, and lattice constant, respectively. The associated {\it sd} exchange coupling energy at the FM interface}
\begin{equation}
\mathcal{F}_{sd}=J_{sd}\frac{M}{M_{s}}\vec{s}\cdot {\vec{m}},
\label{eq:sd}
\end{equation}
\textcolor{black}{where $\vec{m}$ is a unit vector in the direction of  $\vec{M}$.  $M_s$ is the intrinsic saturation magnetization.} Within the Stoner mean-field theory  \cite{Soulen R.J:1998} the spin polarization $\eta$ of the electron density in transition FM metals is usually less than 1, \textcolor{black}{we can decompose the induced spin density $\vec{s}$ as,  \cite{Jia C. L:2014}}
\begin{equation}
\vec{s}=\vec{s}_{\parallel}+\vec{s}_{\perp}
\end{equation}
where $\vec{s}_{\parallel}$ represents the spin density whose direction follows  adiabatically  the intrinsic magnetization $\vec{M}$ at an instantaneous time t. $\vec{s}_{\perp}$ describes the transverse deviation from $\vec{M}$. Given that the steady-state charge accumulation  entails  much higher  energy processes   than  spin excitations, \textcolor{black}{in the absence of a charge current across the FE/FM interface, the spin diffusion normal to the FM/FE interface (hereafter refereed to as the $z$ direction with its origin at the interface) follows the dynamic equation (see Refs.[\onlinecite{Jia C. L:2014},\onlinecite{Manchon:2012jz}] for details)}
\begin{eqnarray}
&&\frac{\partial\vec{s}_{\parallel}}{\partial t}\vec{m}+\vec{s}_{\parallel}\frac{\partial\vec{m}}{\partial t}+\frac{\partial\vec{s}_{\perp}}{\partial t} -D_{0}\nabla^{2}_{z}\vec{s}_{\parallel}-D_{0}\nabla^{2}_{z}\vec{s}_{\perp} \nonumber \\
&&= -\frac{\vec{s}_{\parallel}}{\tau_{sf}}-\frac{\vec{s}_{\perp}}{\tau_{sf}}-\frac{\vec{s}_{\perp}\times\vec{m}}{\tau_{ex}}
\end{eqnarray}
where $D_0$ is the diffusion constant and $\tau_{ex} \approx \hbar/(2J_{sd})$.  $\tau_{sf}$ is the spin-flip relaxation time due to scattering with impurities, electrons, and phonons; $\tau_{sf}\sim 10^{-12}-10^{-14}$ s \cite{L. Piraux:1998} and $\tau_{ex}/\tau_{sf}\sim10^{-2}$ in typical FM metals \cite{J. Bass:2007}. The time-derivative terms $\frac{\partial\vec{s}_{\parallel}}{\partial t}$, $\frac{\partial\vec{m}}{\partial t}$ and $\frac{\partial\vec{s}_{\perp}}{\partial t}$ below THz are negligible compared with $\vec{s}/\tau_{sf}$ and $\vec{s}/\tau_{ex}$.  Thus the steady state is set by \cite{Jia C. L:2014}
\begin{equation}
D_{0}\nabla^{2}_{z}\vec{s}_{\parallel}=\frac{\vec{s}_{\parallel}}{\tau_{sf}}  ~~~\text{and}~~~
D_{0}\nabla^{2}_{z}\vec{s}_{\perp}=\frac{\vec{s}_{\perp}\times\vec{m}}{\tau_{ex}},
\end{equation}
implying an exponentially decaying spiral spin density, \cite{Jia C. L:2014} 
\begin{gather}
\label{eq:14}
s_{\parallel}=\eta\frac{\sigma_{FM}}{\lambda_{m}e}e^{-z/\lambda_{m}}, \\
\vec{s}_{\perp}=(1-\eta)Q_{m}\frac{\sigma_{FM}}{e}e^{-(1-i)\vec{Q}_{m}\cdot\vec{r}}.
\end{gather}
Here $\sigma_{FM} = \sigma_{FE} \approx \epsilon_{FE} E$ is the surface charge density due to the electric neutrality constraint at the interface, $\epsilon_{FE}$  and $E$ are the dielectric permittivity of FE and an applied normal electric field, respectively. $\lambda_{m}=\sqrt{D_{0}\tau_{sf}}$ is the effective spin-diffusion length and the normal spin spiral wave vector $\vec{Q}_{m}=\frac{1}{\sqrt{2D_{0}\tau_{ex}}}\hat{\vec{e}}_{z}$.  Clearly, in the presence of the exchange interaction with long-range FM ordering, the accumulated  (magnonic) spin density extends in the FM system over a {nanometer} characteristic length ($\sim \lambda_m$ being 38 $\pm$ 12 nm in Co \cite{J. Bass:2007})  which is much larger {than} the electrostatic screening length (a few angstroms), albeit both are associated by largely different energy scales.

\begin{figure}[t]
  \centering
  \includegraphics[width=0.5 \textwidth,angle=0]{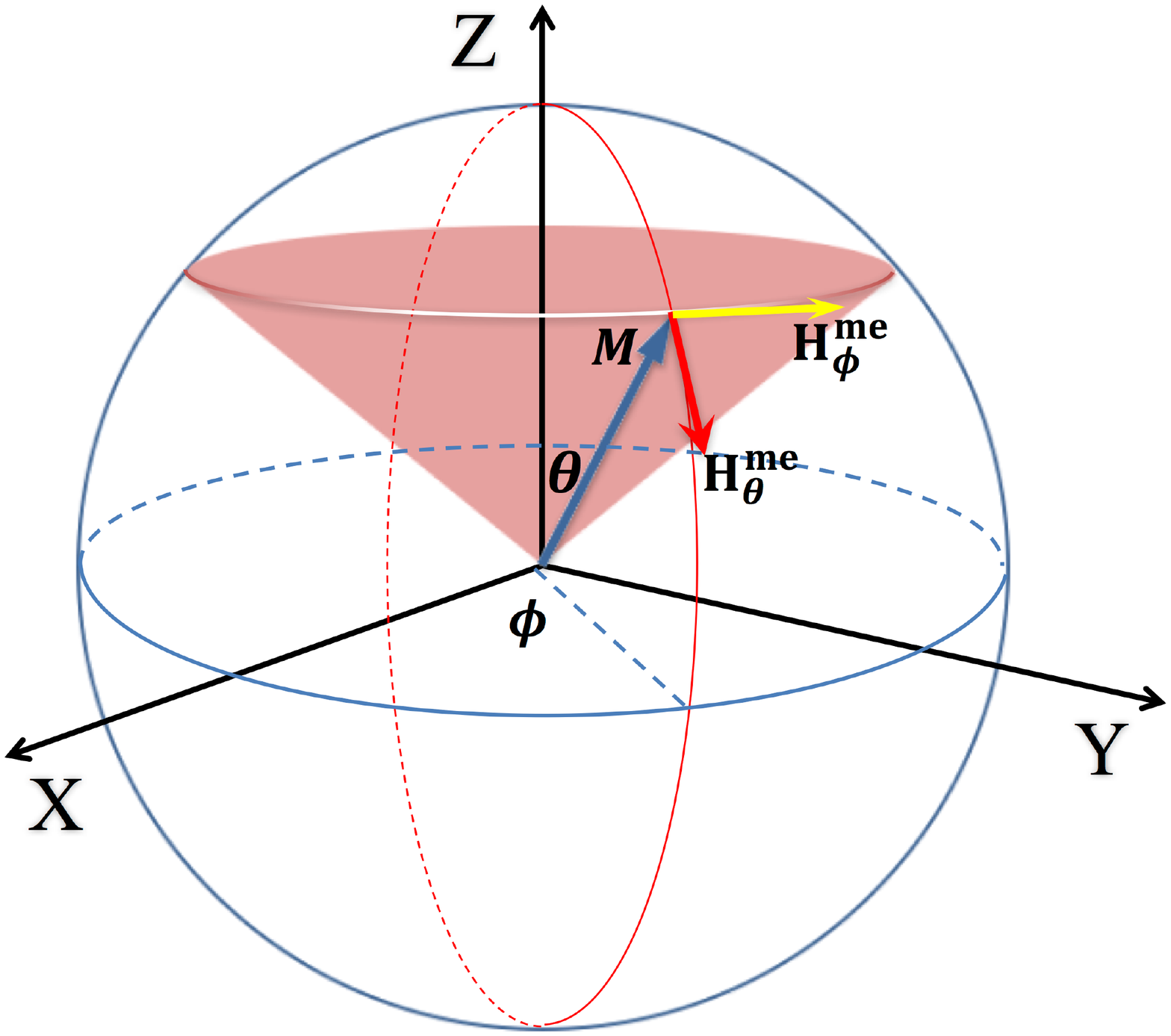}
  \caption{Schematics of the plane of variation for the magnetization $\vec{M}= M \{\cos\phi\sin\theta,\sin\phi\sin\theta,\cos\theta\}$. The FM/FE interface is referred to as the $xy$ plane. $\vec{H}^{\rm{me}}_{\theta}$ and $\vec{H}^{\rm{me}}_{\phi}$ are the transversal components of the interface magnetoelectric field.}
  \label{axis}
\end{figure}

As we {are} interested in the effect of the low-energy accumulated magnonic density on the  spin dynamic in FM  we can safely assume that the spin-dependent charge    excitations are frozen (because of the higher  energy scale) during the  (GHz-THz) spin dynamics in the FM.
\textcolor{black}{Treating the  magnetic dynamics  we consider the additional effective magnetoelectric field $\vec{H}^{\rm{me}}$ acting on  the magnetization dynamics $\vec{M}(t)$ due to the interfacial   spin order. To leading terms, from  the {\it sd} interaction energy [Eq. (\ref{eq:sd})] we derive}
\begin{equation}
\vec{H}^{\rm{me}}= - \delta \mathcal{F}_{sd}/\delta{\vec{M}} = -\frac{J_{\text{sd}}}{M_{s}}\vec{s}.
\end{equation}
We choose nanometer thick layers Co and BaTiO$_3$ as   prototypical  FM and FE layers for estimating the  characteristics of $\vec{H}^{\rm me}$. The density of surface charges \cite{J. Hlinka:2006} reads $\sigma_{\text{FE}} =  0.27$ C/$m^{2}$ and the parameters of Co are \cite{J. M. D. Coey:2010}: $M_{s}=1.44\times10^{6}$ A/m, $K_{1}=4.1\times 10^{5}$ J/$m^{3}$, $\lambda_{m}= 40$ nm \cite{J. Bass:2007}, and $\eta=0.45$ \cite{Soulen R.J:1998}. We find thus $|\vec{H}^{\rm{me}}| \approx 0.2$ T with {$J_{sd} \approx 0.1$ eV/atom} and the FM thickness $d_{FM} = 40$ nm.  Such a strong magnetoelectric field is comparable with the uniaxial anisotropic field $\frac{K_{1}}{M_{s}} \approx 0.3$ T of Co. More importantly, note that the non-adiabatical component $\vec{H}^{\rm{me}}_{\perp}$ is always perpendicular to the direction of magnetization $\vec{M}$, acting as a  field-like torque and a damping-like torque at all time (c.f. Fig.\ref{axis}), which would play a key role  for  electric-field assisted magnetization switching.

\textcolor{black}{\section{Magnetization dynamics}\label{sec2}}

We start from the Landau-Lifshitz-Baryakhtar equation (LLBar) \cite{V.G.Baryakhtar:1984,M.Dvornik:2013,Weiwei Wang:2015} for the magnetization dynamics at the FM interface,
\begin{equation}
\label{eq:1}
\frac{\partial\vec{M}}{\partial t}=-\gamma\vec{M}\times\vec{H}_{\text{eff}}+\hat{\Lambda}_{r}\cdot\vec{H}_{\text{eff}}-\hat{\Lambda}_{e,ij}\frac{\partial^{2}\vec{H}_{\text{eff}}}{\partial x_{i}\partial x_{j}}
\end{equation}
where $\gamma$ is the gyromagnetic ratio.  The last two terms describe the local and nonlocal relaxations. $\hat{\Lambda}_{r}$ and $\hat{\Lambda}_{e}$
 are generally the relaxation tensors of relativistic and exchange natures, respectively. \textcolor{black}{The anisotropy of relaxations decreases with
increasing temperature. Experimentally isotropy of relaxations  were discussed in Ref.[\onlinecite{Seib:2009es}]. We can represent the relaxation tensors as $\hat{\Lambda}_{r}=\lambda_{r}$ and $\hat{\Lambda}_{e}=\lambda_{e}$ where $\lambda_{r}=\gamma\alpha M_{s}$ and $\lambda_{e}={\gamma g\mu_{B}\hbar G_{0}}/({8e^{2}})$ with $\alpha$ and $G_{0}$ being the Gilbert damping coefficient and the conductivity of FM system, respectively.
$e$ is the electron charge.} In contrast to the Landau-Lifshitz-Gilbert equation, the LLBar equation does not conserve the magnitude of the magnetization capturing  the magnetic relaxations in metals, especially the case for FM metal interfaces. This is necessary in our case to ensure that
 the local magnetic order which is
in equilibrium with the interface region relaxes to the asymptotic bulk magnetization.

By introducing $\vec{M} = M {\vec m}$ into the LLBar equation, we infer the following equation for the direction of magnetization \cite{Weiwei Wang:2015},
\begin{equation}
\label{eq:6}
\frac{\partial\vec{m}}{\partial t}=-\gamma\vec{m}\times\vec{H}_{eff}+\frac{1}{M_{s}}\vec{R}_{\perp}
\end{equation}
with $\vec{R} = \lambda_{r}\vec{H}_{eff}-\lambda_{e}\nabla^{2}_z\vec{H}_{eff}$ and $\vec{R}_{\perp}=-\vec{m}\times(\vec{m}\times\vec{R}).$
 Here
  \begin{equation}
   \vec{H}_{eff} = \vec{H}_{eff}^0 + \vec{H}^{\rm me}
  \end{equation}
  is the effective magnetic field, in which $\vec H_{eff}^0 $ follows from the functional derivative of the
  free energy density via  \cite{Sukhov:2014}
\begin{eqnarray}
\vec{H}_{eff}^0& =& -\delta \mathcal{F}_{0}/\delta \vec{M}, \nonumber \\
 \mathcal{F}_{0}&=&-K_{1}(\sin^{2}\theta\cos^{2}\phi\sin^{2}\theta_{u}+\cos^{2}\theta\cos^{2}\theta_{u}) \nonumber \\
 & &-\frac{K_{1}}{2}\sin2\theta\sin2\theta_{u}\cos\phi-(K_{s}/{d_{FM}}-\mu_{0}M^{2}_{s}/2)\cos^{2}\theta \nonumber \\
 & &-\vec{M}\cdot\vec{B}.
 \label{eq:F0}
\end{eqnarray}
 $K_{1}$ is the  uniaxial magnetocrystalline anisotropy energy, $K_{s}$ is the magnetic surface anisotropy contribution which is significant for
 relatively thin magnetic film and favors magnetization out of the $xy$ plane.
 $\mu_{0}M^{2}_{s}$ denotes the demagnetizing field contribution, which favors a magnetization in plane.
  $\vec{M}\cdot\vec{B}$ is the Zeemann interaction and $\theta_{u}$ is the tilted angle of the easy axis  from the $z$ direction.

Clearly, the non-uniform effective field $ \vec{H}^{\rm me}$ due to the \emph{s-d} interaction with the exponentially decaying
spiral spins would give rise to  nonlocal damping of the magnetization dynamics.
Considering that the contribution of the induced spin density to {the spatial distribution of} local ferromagnetic moments is small, we have
\begin{equation}
\langle \nabla^{2}_{z}\vec{s}_{\perp}\rangle=2Q^{2}_{m}\left( \langle s^{\phi}_{\perp}\rangle \hat{\vec e}_{\theta}-\langle s^{\theta}_{\perp}\rangle \hat{\vec e}_{\phi} \right).
\end{equation}
Without loss of generality one can take $\langle s^{\phi}_{\perp}\rangle =\langle s^{\theta}_{\perp}\rangle =\frac{1}{\sqrt{2}}\langle s_{\perp}\rangle $. It is also convenient to redefine some dimensionless parameters which are $\tilde{d}_{FM}=\frac{d_{FM}}{\lambda_{m}}$, $\tilde{t}$= t$\gamma$T$\approx$ 28t GHz, and $\tilde{J}_{sd}=\frac{J_{sd}}{\rm eV}\frac{\sigma_{FM}}{P_{s}}\frac{\lambda_{m}}{d_{FM}}$ with the FE spontaneous polarization $P_s$.  In the following $\tilde{J}_{sd}$ is taken as an adjustable parameter in view of ferroelectric tuning of magnetoelectric field $ \vec{H}^{\rm me}$.

\textcolor{black}{\section{Numerical results and discussions}\label{sec3}}

For the surface anisotropy $K_{s}\approx10^{-3}$ J/$m^{2}$ and $\mu_{0}M^{2}_{s}/2 \approx$ 1.3$\times10^{6}$ J/$m^{3}$ of Co sample \cite{J. M. D. Coey:2010},  the dominant contribution of the anisotropic term $(K_{s}/{d_{FM}}-\mu_{0}M^{2}_{s}/2)$  in Eq.(\ref{eq:F0}) has the form either $K_{s}/{d_{FM}}$, or $-\mu_{0}M^{2}_{s}/2$ depending on the thickness $d_{FM}$, i.e., the magnetization will be either normal to the FM interface ($\theta_u =0$) or in the interface plane ($\theta_u = \pi/2$).

\emph{Case I}: Normal FM magnetization with $\theta_u =0$:  The free energy density is
\begin{equation} \mathcal{F}_{0}=-K_{eff}\cos^{2}\theta-\vec{M}\cdot\vec{B},\quad
 K_{eff}= K_1 + \frac{K_s}{d_{FM}} - \frac{\mu_0 M_s^2}{2}\end{equation}
which leads to  \begin{equation} \vec{H}^{0}_{eff}=\frac{2K_{eff}}{M_{s}}\cos\theta \hat{\vec e}_z \end{equation} without an applied magnetic field $\vec{B}$. The LLBar equation reads then,
\begin{eqnarray}
& \frac{\partial\theta}{\partial \tilde{t}}=\frac{\gamma^{e}_{+}}{\sqrt{2}} {H}^{\rm me}_{\perp}-\alpha \frac{K_{eff}}{M_{s}}\sin2\theta,
\\
&\sin\theta\frac{\partial\phi}{\partial \tilde{t}}=-\frac{\gamma^{e}_{-}}{\sqrt{2}} {H}^{\rm me}_{\perp}+\frac{K_{eff}}{M_{s}}\sin2\theta
\end{eqnarray}
{with } $\gamma^{e}_{\pm}=1 - 2Q^{2}_{m}\frac{\lambda_{e}}{\gamma M_{s}}\pm \frac{\lambda_{r}}{\gamma M_{s}}= \gamma^{e}\pm \alpha$  .

\begin{figure}[t]
  \centering
  \includegraphics[width=0.9 \textwidth,angle=0]{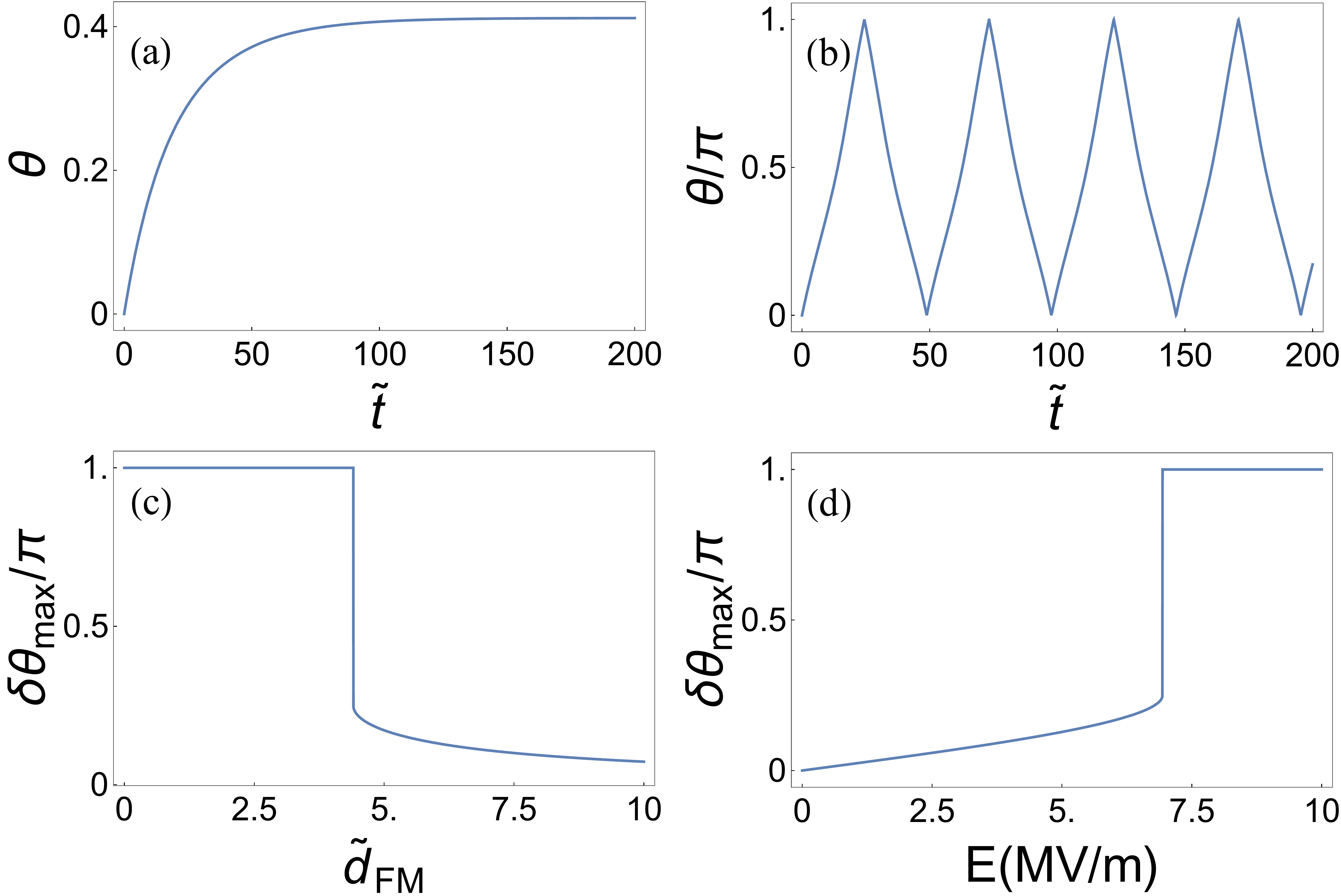}
  \caption{Dynamics of the normal magnetization. The polar angle $\theta$ vs. dimensionless time $\tilde{t}$ for different ME field (a) $\tilde{J}_{sd}=0.005$ and (b) $\tilde{J}_{sd}=0.03$, respectively. Panels (c) and (d) demonstrate the thickness and electric-field dependence of $\delta \theta_{max} = \theta_{max}-\theta(0)$, where $\theta(0)$ and $\theta_{max}$ are respectively the initial value ($\theta(0)=0$) and the maximum value of the polar angle during the time  evolution of magnetization.   $\delta \theta_{max} = \pi$ indicates a magnetization flip over the normal $\hat{\vec e}_z$ direction.   Here, $\alpha = 0.1$, $\epsilon_{FE} =1000$ and {$K_{eff}\sim K_{1}$}.}
  \label{fig2}
\end{figure}

 Clearly, under a weak interfacial ME field, the condition \begin{equation}{H}^{\rm me}_{\perp} = \sqrt{2} \frac{\alpha}{\gamma^{e}_{+}} \frac{K_{eff}}{M_{s}}\sin2\theta \end{equation} can be satisfied,  the polar angle $\theta$  ends up processionally in the equilibrium state [c.f. Fig.\ref{fig2}(a) with
 ${\partial\theta}/{\partial \tilde{t}}=0$]. Otherwise, the strong transversal field ${H}^{\rm me}_{\perp}$  results in  a magnetization flip
 over the normal $\hat{\vec e}_z$-direction [Fig.\ref{fig2}(b)].
 Considering that the ME field depends linearly on the applied electric field and the reciprocal of  FM thickness, one would expect a transition from the magnetization procession around the z axis (for a small electric field $E$ and/or relatively thick FM layers) to the magnetization {flip over} the normal {direction} (for a strong electric field and/or ultra-thin FM film) at the critical points, as demonstrated in Figs. \ref{fig2}(c) and \ref{fig2}(d).

\begin{figure}[t]
  \centering
  \includegraphics[width=1 \textwidth, angle=0]{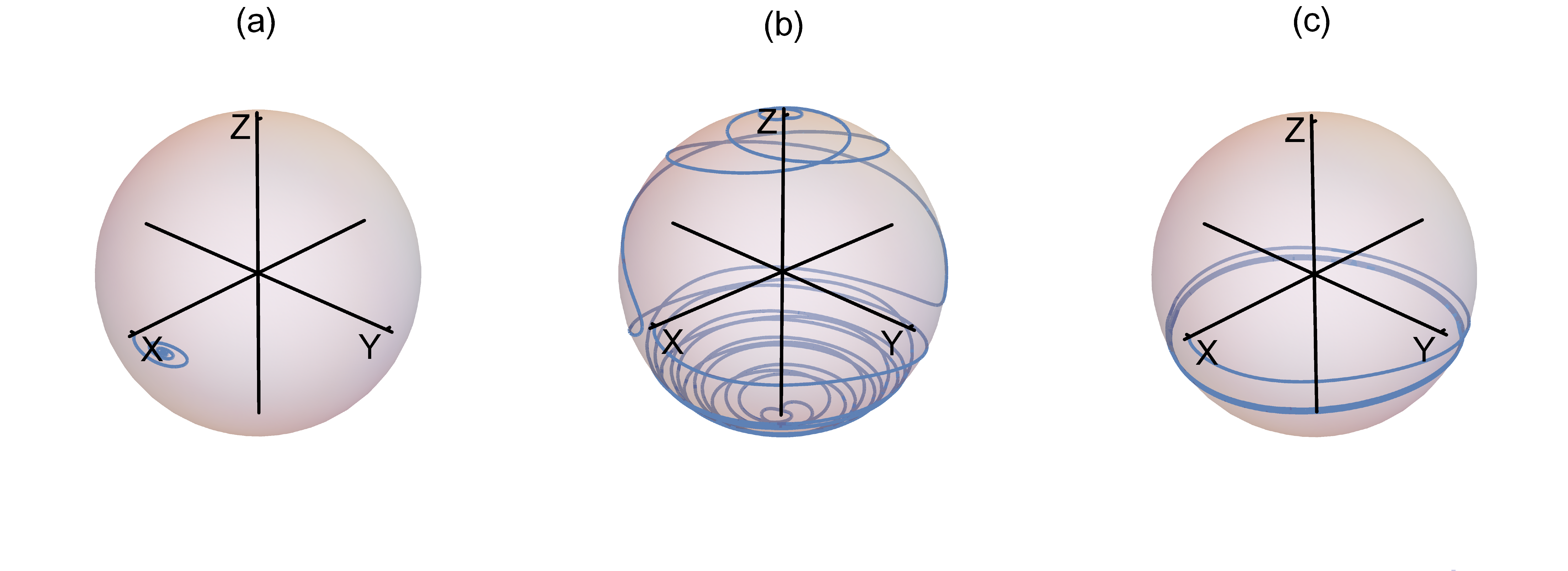}
  \caption{Dynamics of the in-plane magnetization for different interface ME field and anisotropic field: (a) {${\tilde{J}}_{sd}=0.03$} and $2K_1/M_s=0.6$ T, (b) {${\tilde{J}}_{sd}=0.03$} and $2K_1/M_s=0.3$ T, and (c) {${\tilde{J}}_{sd}=0.015$} and $2K_1/M_s=0.1$ T, respectively. Here {${\tilde{d}}_{FM}=1$} and $\alpha=0.1$.}
  \label{f3}
\end{figure}
\begin{figure}[t]
  \centering
  \includegraphics[width=0.7 \textwidth,angle=-90]{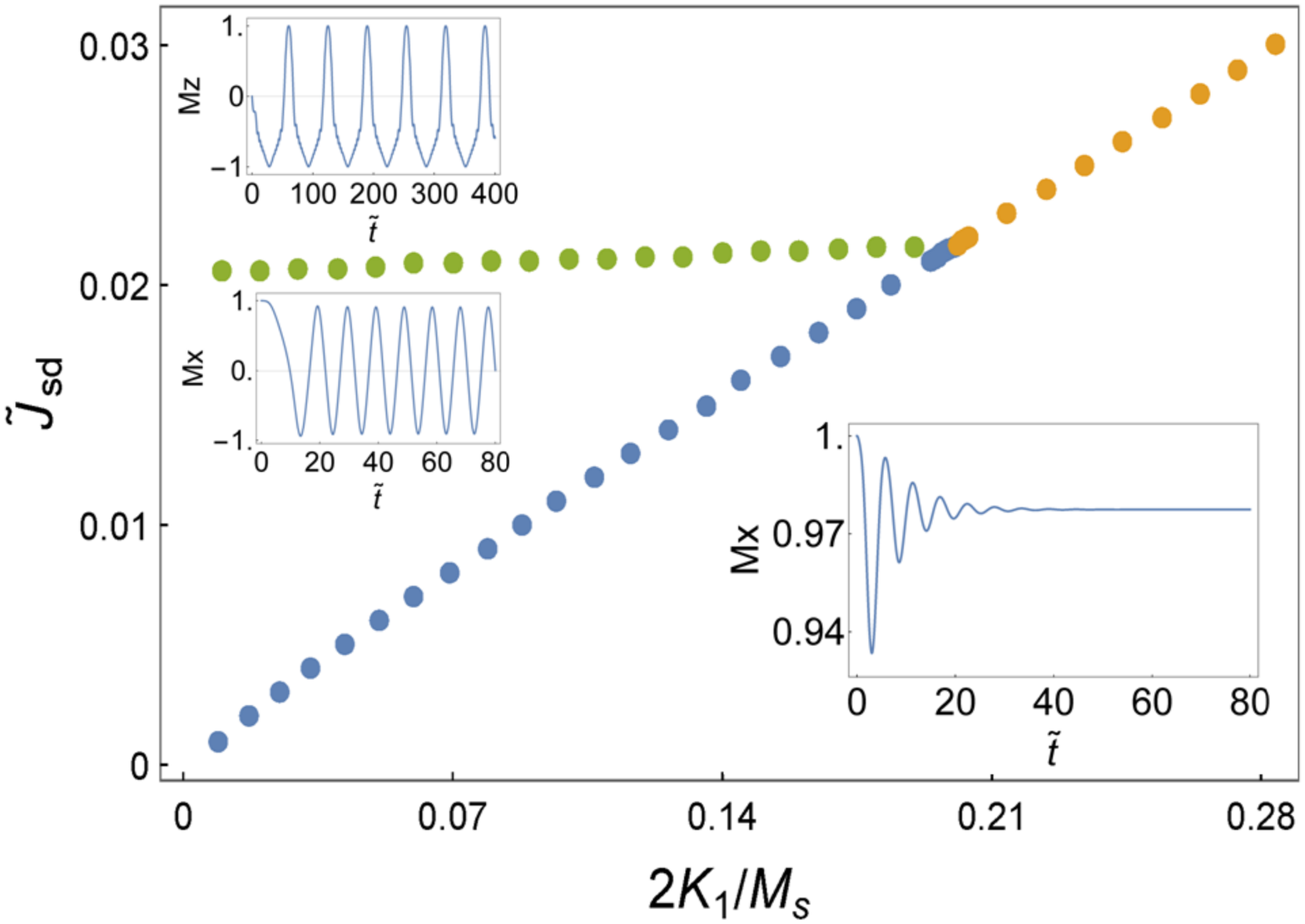}
  \caption{\textcolor{black}{Phase diagrams of the in-plane magnetization dynamics with $\alpha=0.1$, $\tilde{d}_{FM}=1$, and $\vec{B} =0 $: (a) the localized precessional mode, (b) the $z$-axial flip mode, and (c) the $z$-axial rotational mode, respectively. The characterization of dynamic behavior of magnetization in three different phases is illustrated in Fig.\ref{f3}. Insets show the corresponding time evolution of magnetization in each phase.}}
  \label{fig7A1}
\end{figure}

\emph{Case II}: In-plane magnetization with $\theta_u = \pi/2$: Disregarding the surface anisotropy ($K_{s}/d_{FM} \ll \mu_{0}M^{2}_{s}/2$) for a thick FM film, the effective magnetic field reads
\begin{equation}
\vec{H}^{0}_{eff}=2K_{1}/M_{s} \sin\theta\cos\phi \hat{\vec e}_{x} -\mu_{0}M_{s}\cos\theta\hat{\vec e}_{z}+\vec{B}
\end{equation}
and the magnetization favors an in-plane $\hat{\vec e}_x$ axis, which means $\phi(0) =0$ with the external magnetic field $\vec B =0$.
Upon some simplifications the LLBar equation reads

\begin{eqnarray}
\frac{\partial\theta}{\partial \tilde{t}}&=&\frac{\gamma^{e}_{+}}{\sqrt{2}}{H}^{\rm me}_{\perp}+\alpha \frac{\mu_{0}M_{s}}{2}\sin2\theta + \alpha \frac{K_{1}}{M_{s}}\sin2\theta\cos^{2}\phi  \nonumber \\
&-& \frac{K_{1}}{M_{s}}\sin\theta\sin2\phi + \alpha B\cos\theta\cos\phi- B\sin\phi,
\\
\sin\theta\frac{\partial\phi}{\partial \tilde{t}}&=&-\frac{\gamma^{e}_{-}}{\sqrt{2}}{H}^{\rm me}_{\perp}-\frac{\mu_{0}M_{s}}{2}\sin2\theta - \alpha \frac{K_{1}}{M_{s}}\sin\theta\sin2\phi \nonumber \\
&-& \frac{K_{1}}{M_{s}}\sin2\theta\cos^{2}\phi -B\cos\theta\cos\phi- \alpha B\sin\phi.
\end{eqnarray}
In the absence of external magnetic field $\vec B$, the magnetization dynamics is determined by three parameters: $\alpha$, $H^{\rm me}_{\perp}$, and $K_1/M_s$. Firstly, let us ignore the damping terms for small Gilbert damping coefficient $\alpha$, the weak ME field $H^{\rm me}_{\perp}$ would satisfy $\frac{\partial\theta}{\partial \tilde{t}}=0$ and $\frac{\partial\phi}{\partial \tilde{t}} =0$, resulting in a relocation of the magnetization with an equilibrium tilted angle {in the vicinity of $x$ axis}, as shown in Fig.\ref{f3}(a). However, when $H^{\rm me}_{\perp}$ is  stronger  than the anisotropic field $K_1/M_s$ and the demagnetization field $\mu_0 M_s$, no solutions exist for $\partial \theta/\partial \tilde{t} =0$ at all time, the magnetization possesses {a $z$}-axial {flip} mode in the {whole spin space} [c.f. Fig.\ref{f3}(b)] similar to the case of normal FM magnetization. On the other hand, after accounting for terms containing $\alpha$ in the LLBar equations, we would have additional magnetization rotation around the $z$ axis [Fig.\ref{f3}(c)]. Further insight into the detailed characterization of magnetization dynamics is delivered by numerics  for a varying strength of the ME field $\tilde{J}_{sd}$ and the uniaxial anisotropy $K_1/M_s$ in Fig.\ref{fig7A1} with $\alpha=0.1$. There are two new phases, the {$z$-axial flip} mode and the $z$-axial rotational mode, which were unobserved in the FM systems in the absence of interface ME interaction. With decreasing the damping $\alpha$, the area of the $z$-axial rotational mode  shrinks vanishing eventually. By applying an external magnetic field $\vec B$ along the $x$-direction, only slight modifications are found in the phase diagram. However, the initial azimuthal angle $\phi(0)$  deviates from the easy axis with a rotating magnetic field $\vec B$ in the $xy$ interface plane. Considering the LLBar equations with the initial condition $\theta(t=0) = \pi/2$, we have
\begin{equation}
      \frac{\partial\theta}{\partial \tilde{t}}|_{\tilde{t}=0}
       \approx \frac{\gamma^{e}_{+}}{\sqrt{2}}\vec{H}^{s-d}_{\perp}-\frac{K_{1}}{M_{s}}\sin2\phi(0)
\end{equation}
with a small damping $\alpha$.
       As the dynamic equation is sensitive to the initial azimuthal angle $\phi(0)$, the calculations show that the magnetization dynamics may change between the processional mode around the $x$ axis and the {$z$-axial flip or $z$-axial rotational} mode, depending on the initial value of $\phi(0)$.

Phenomenologically, such an $z$-axial flip mode and  $z$-axial rotational mode are exhibited  as a precessional motion of the magnetization with a negative damping,
as shown in the experimental observation for polycrystalline CoZr/plumbum magnesium niobate-plumbum titanate (PMN-PT) heterostructures \cite{Jia:2015iz},
where an emergence of positive-to-negative transition of magnetic permeability was observed by applying external electric field.
There is also some analogy between these non-equilibrium switching behaviors in FM/FE heterostructures and the negative
damping phenomenon in trilayer FM/normal metal/FM structures, in which the supplying energy is thought to be provided
by injecting spin-polarized electrons from an adjacent FM layer, magnetized in the opposite direction compared to the FM layer under consideration \cite{Shufeng Zhang:2009,M-Book}.

\begin{figure}[b]
  \centering
  \includegraphics[width=0.65 \textwidth,angle=0]{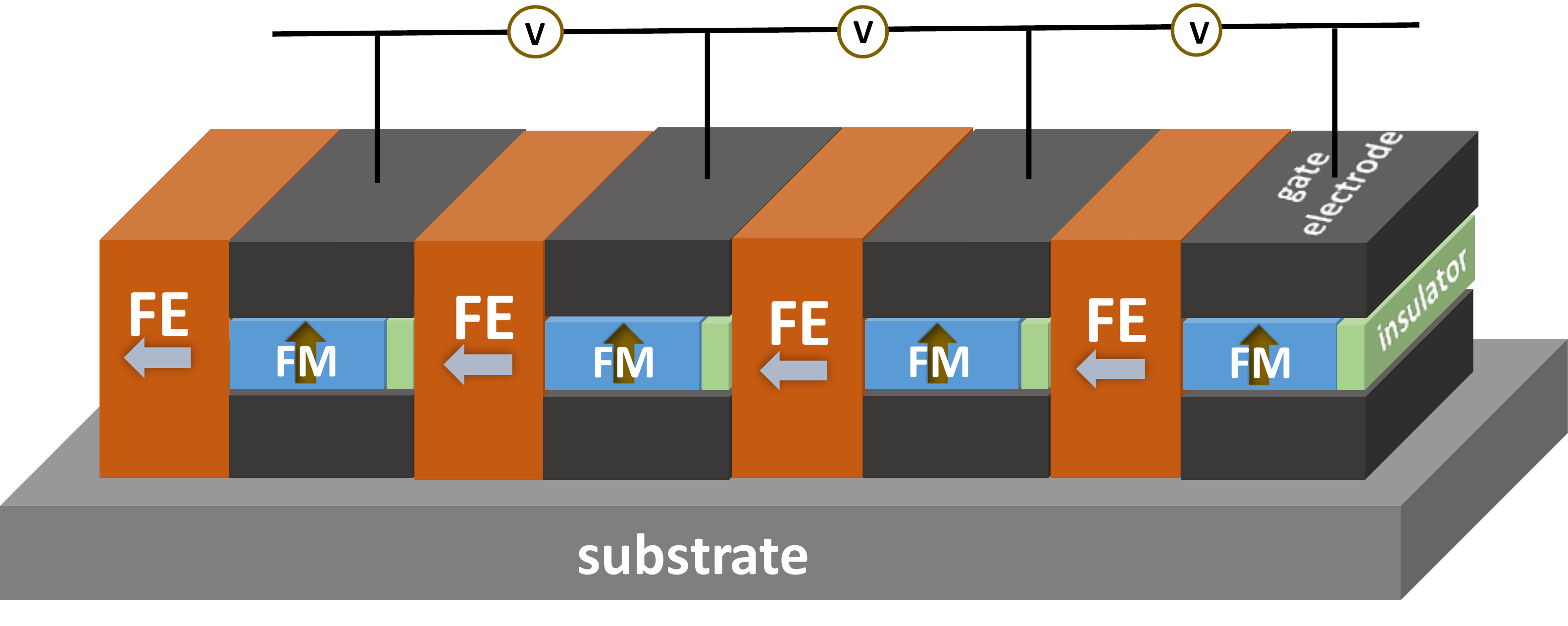}
  \caption{\textcolor{black}{Schematic structure diagram of the FE/FM multilayer system with enhanced ME effects. The arrows mark the directions of the FE polarization $\vec{P}$ and the FM magnetization $\vec{M}$, respectively.}}
  \label{fig5}
\end{figure}

\textcolor{black}{\section{Conclusion and outlook}\label{sec4}
The above theoretical considerations along with numerical simulations for specific material FE/FM composites endorse that
the magnetization dynamics can be controlled by an electric field of a moderate strength.   The excitations triggered by the electric field are transferred to the spin system via the interface spiral-mediated magnetoelectric  coupling  and may result in a magnetization switching.  This direct electric field control of magnetization switching offers a {qualitatively} different way to manipulate magnetic devices swiftly  with  low-power write capability. On the other hand, even though  the spin-mediated magnetoelectric coupling has a much longer range that the surface localized charge-mediated FE/FM coupling, its range is still limited by the spin diffusion length which
is material dependent but yet in the range of several tens of nanometers. Hence, the full power of the predicted effect is expected for multilayer systems such as those schematically shown in Fig.\ref{fig5}: Starting from a bilayer structure with a thick FE interfaced with a FM layer which has a  thickness in the range of the spin diffusion length, we suggest to cap this structure with
  a spacer layer, for instance an (oxide) insulator. Repeating the whole structure as proposed in Fig.\ref{fig5} allows for a simple serial extension  from a double to multilayer structure while enhancing the influence of the magnetoelectric coupling.}

\textcolor{black}{\section{Acknowledgment}}
{
This work is supported the National Natural Science Foundation of China (Grant No. 11474138), the German Research Foundation (Grant No. SFB 762), the Program for Changjiang Scholars and Innovative Research Team in University (Grant No. IRT-16R35), and the Fundamental Research Funds for the Central Universities.
}


\end{document}